# Two-XUV-photon double ionization of Neon studied at the Extreme Light Infrastructure (ELI-ALPS)

I. Orfanos[1], E. Skantzakis[1], A. Nayak[2,3], M. Dumergue[2], S. Kühn[2], G. Sansone[2,4], M.F. Kling[5,6,7,8], H. Schröder[5,6], B. Bergues[5,6], S. Kahaly[2], K. Varju[2], A. Forembski[9], L.A.A. Nikolopoulos[9], P. Tzallas[1,2], D. Charalambidis[1,2,10*]

[1]*Foundation for Research and Technology - Hellas, Institute of Electronic Structure & Laser, PO Box 1527, GR71110 Heraklion (Crete), Greece.*

[2] *ELI-ALPS, ELI-HU Non-Profit Ltd., Wolfgang Sandner utca 3., H-6728 Szeged, Hungary.*

[3]*University of Szeged, Dugonics tér 13, H-6720 Szeged, Hungary.*

[4]*Physikalisches Institut, Albert-Ludwigs-Universität, Stefan-Meier-Straße 19, 79104 Freiburg, Germany.*

[5]*Department of Physics, Ludwig-Maximilians-Universität Munich, D-85748 Garching, Germany.*

[6]*Max-Planck-Institut für Quantenoptik, Hans-Kopfermann-Strasse 1, D-85748 Garching, Germany.*

[7]*SLAC National Accelerator Laboratory, Menlo Park, CA 94025, USA.*

[8]*Applied Physics Department, Stanford University, Stanford, CA 94305, USA.*

[9]*School of Physical Sciences, Dublin City University, Collins Ave, Dublin 9, Ireland.*

[10]*Department of Physics, University of Crete, PO Box 2208, GR71003 Heraklion (Crete), Greece.*

[*] Corresponding author e-mail address: chara@iesl.forth.gr

## Abstract

Two XUV-photon double ionization of Ne, induced by an intense few-pulse attosecond train with a ~ 4 fs envelope duration is investigated experimentally and theoretically. The experiment is performed at ELI-ALPS utilizing the recently constructed 10 Hz gas phase high-order harmonic generation SYLOS GHHG-COMPACT beamline. A total pulse energy up to ~1 μJ generated in Argon in conjunction with high reflectivity optics in the XUV region, allowed the observation of the doubly charged state of Ne induced by 40 eV central XUV photon energies. The interaction of the intense attosecond pulse train with Ne is also theoretically studied via a second-order time dependent perturbation theory equations-of-motion. The results of this work, combined with the feasibility of conducting XUV-pump-XUV-probe experiments, constitute a powerful tool for many potential applications. Those include attosecond pulse metrology as well as time resolved investigations of the dynamics underlying direct and sequential double ionization and their electron correlation effects.

**Introduction**

When an atomic/molecular system is exposed to an external electromagnetic field a wide variety of processes may occur, depending on the interaction strength and wavelength of the radiation. At low intensities linear processes are occurring which can be described by first-order time-dependent perturbation theory. At increased intensities, typically $> 10^{10}$ W/cm$^2$, nonlinear processes [1,2] start playing a significant role, as theoretically predicted [1] and experimentally inaugurated in the 60s [2,3]. Here the wavelength of the radiation is a decisive parameter whether the interaction is multi-photon, commonly treated through lowest order time dependent perturbation theory (LOPT), or strong field (non-perturbative) type [4-7]. Besides higher order harmonic generation (HHG) [8-12], which possess the most exciting applications, of particular interest among these nonlinear processes is the double (or multiple) ionization. Double ionization can be a direct (without formation of intermediate ions) or sequential (with formation of lower charge states that ionize further) process. The particular interest is due to the possible electron-electron correlation effects and their dynamics that may underlay the process.

Although extensively explored in the visible and infrared (IR) regions, the extension of the studies of such processes to the extreme ultraviolet (XUV) is limited mostly to accelerator-based facilities like Free Electron Lasers (FEL) [13-15] (and references therein), offering remarkably high pulse energies, but with shot-to-shot instabilities [16,17] and, until recently [18,19], relatively long pulse durations. On the other hand, coherent, table-top, laser-driven sources based on gas phase HHG are versatile, provide routinely sub-fs temporal resolution and have relative good shot to shot stability. The inherently low conversion efficiency of the generation process, however, limits the number of the emitted photons. To this end, extended efforts in several laboratories led to the development of novel HHG beamlines that have reached pulse energies and durations that allowed the observation of non-linear XUV processes [20-30], recently even beyond the two-XUV-photon ionization [31-34], as well as strong-XUV-field effects [35]. Inducing such phenomena is a vital tool in studying ultrafast dynamics, that can be used to initiate, probe or even control processes that are occurring on the few femtosecond (fs) – attosecond (as) timescales in atomic and molecular systems [36-44] (and references therein). The key feature in intense coherent XUV sources, based on HHG in gaseous media, is loose focusing. This allows the interaction with high laser pulse energies without depleting the generation medium. Loose focusing geometries are exploited in the so called "SYLOS GHHG-COMPACT beamline" at the ELI-ALPS user research infrastructure [45]. With this beamline high order harmonic emission with pulse energies in the μJ range has recently been achieved. These pulse energies in conjunction with very short pulse durations were evaluated as sufficient for inducing nonlinear processes in the XUV region.

In this work, we investigate the two-photon double ionization of Ne induced by intense XUV radiation. The doubly charged Ne species are produced by a few-pulse train with an estimated envelope pulse duration of 3 - 4 fs, synthesized by a comb of harmonics ($23^{rd}$-$31^{st}$) centered at 40 eV photon energy and intensity of the order of $10^{12}$ W/cm$^2$. The experimental results are supported by theoretical calculations that evaluate the possible double ionization pathways of Ne, indicating that both the direct and the sequential processes are significantly contributing to the double ionization. Thus, we observe a two-XUV-photon double ionization, with significant contribution of the direct process, while the two photon sequential channel is partially open. To our knowledge these are the highest harmonic comb photon energies that have induce so far such an ionization scheme. The present results establish the SYLOS GHHG-COMPACT beamline of ELI-ALPS as an intense attosecond [46] XUV beamline with which users can study or exploit nonlinear processes in the XUV spectral region. The work presented in this paper is aiming to disseminate the availability of such a user beamline and its capacity to the international community. While here the 10Hz SYLOS system has been used, similar results will in the near future be produced with the 1kHz SYLOS laser chain [47], enabling for the first time electron-electron and ion-electron coincidence XUV-pump-XUV-probe experiments with sub-fs temporal resolution.

**Experimental methods**

In the present work the attosecond beamline is driven by the 10 Hz stage of the TW class SYLOS Laser system, delivering IR pulses with energy up to 40 mJ and 10 fs duration at a carrier wavelength of 850 nm [48]. Based on the concept of loose-focusing of the driving field, the compressed IR radiation is steered towards the interaction region by means of a 10 m focal length spherical mirror. The long focusing configuration, as discussed above, allows the interaction with high laser pulse energies, increasing at the same time the number of the emitters contributing to the total XUV emission [49-51]. As a consequence, the beamline generates XUV pulses in the μJ energy range. In particular, when Ar is used as generating medium, the total XUV pulse energy in the generation region is ~1 μJ with the photon energies spanning in the range of 16-50 eV. Although using Xe would have led to higher XUV pulse energies, Ar gas was chosen due to its higher cut-off photon energies, which are more favorable for the excitation of the processes investigated in the present work. Based on recently developed arrangements under similar experimental conditions [31,32], the XUV radiation is generated by implementing a dual pulsed gas jet configuration placed near the focal area of the IR beam. Both gas jets are operated by piezoelectrically driven pulsed nozzles with the backing pressure optimized and set at 3.5 bars. The dual gas jet arrangement, exploiting a quasi-phase matching strategy, is leading to an enhancement compared to a single gas jet. Placing the first gas jet at the focusing plane of the driving field, the emitted XUV energy is optimized by adjusting the pressure. Then the second gas jet is switched on, and by moving it along the propagation axis, a relative enhancement of ~2 is observed at a ~ 14 cm distance between the two jets. The generated XUV radiation and the fundamental IR co-propagate and impinge on a Silicon plate (Si) placed at Brewster's angle (~ 75 deg.) for the IR radiation. The Si plate significantly reduces the amount of the IR radiation and reflects 50-60% of the XUV beam's energy [52]. Additionally, an aperture of 5 mm diameter is placed before the Si reflection, blocking the outer part of the residual IR beam. Furthermore, a mount hosting metallic filters is introduced in the beam path for the spectral selection of the XUV radiation and also eliminating any remaining amount of the fundamental frequency. For the experimental investigations in this work a 150 nm Al filter transmitting the harmonics with order $q \geq 11$ was used. The XUV pulse energy was measured with a calibrated XUV photodiode, placed after the metallic filter. The photodiode was mounted on a translation stage that could move it in and out of the beam path. Moreover, the filtered XUV beam could be spatially characterized by means of an XUV beam profiler, consisting of a pair of multichannel plates (MCPs) and a phosphor screen followed by a CCD camera. Finally, the spectral components of the generated XUV radiation are recorded by a Flat Field Spectrometer (FFS) which is attached to the beam line. After the XUV emission optimization and characterization, the beam is introduced into the detection chamber in which the interaction of the harmonic radiation with the Ne gas is studied. The XUV beam is impinging at almost normal incidence onto a multilayer spherical mirror of 5 cm focal length mounted on a multiple-translation-rotation stage. The multilayer mirror has ~ 40% reflectivity at 40 eV central photon energy with a bandwidth of ± 2 eV and focuses the radiation into the Ne gas target. The Ne gas is supplied into the interaction region by means of a piezo-based pulsed nozzle having its orifice very close to the XUV focus. The interaction products are recorded with a time-of-flight (TOF) ion mass spectrometer.

Taking into account the measured XUV energy, the reflectivity of the Si plate, the transmission of the Al filter as well as the reflectivity of the multilayer XUV focusing mirror, the energy of the harmonic radiation in the interaction region can be estimated. Furthermore, with an evaluation of the XUV focal spot, based on previous experimental investigations with similar XUV focusing geometries **[28]**, as well as of the time duration of the APT's envelope, the maximum peak intensity in the interaction region is estimated to be of the order of $10^{12}$ W/cm$^2$. Although the values reported in this work are expected to be improved as further optimization is still ongoing, already at such intensity levels nonlinear processes are expected to be observed upon interaction with atomic/molecular systems. In fact, excitation of Ne with XUV radiation of 40 eV photon energy, leads to the observation of Ne$^{2+}$ ions via a nonlinear ionization process. In Figure 2 (b) the excitation scheme of Ne is presented, depicting the possible XUV-photon induced double ionization channels for the harmonics transmitted by the Al filter and reflected by the multilayer spherical mirror. Under the influence of the XUV radiation used in the experiment, there are three possible channels producing Ne$^{2+}$ ions. As depicted, these channels are either direct, through simultaneous absorption of two XUV photons of 40 eV energy, or sequential. In the latter case, however, considering the 40 eV central photon energy of excitation, double ionization is requiring a three-photon absorption process. In particular, the first photon absorbed by Ne is leading to the formation of an ionic $Ne^+$ state. Absorption of a second photon may then occur leading to ionization of the $Ne^+$ ion. Finally, contribution of a sequential two-photon double ionization process has to be considered since a small fraction of the ionizing spectrum has photon energies exceeding the ionization potential of $Ne^+$. Indeed, as indicated in the theoretical part of this work, both the sequential and the direct process are considerably contributing to the double ionization under the present experimental conditions. In Figure 2(b) the black field curve depicts a typical measured HHG spectrum produced in Ar, recorded by means of the FFS. The photon energies reach up to ~ 48 eV corresponding to the 33$^{rd}$ harmonic of the frequency of the driving field. While the cut-off of the emitted spectrum is ~100 eV **[46]**, the ~50eV highest measured photon energy results from the Si reflectivity edge **[52]**. The red dashed curve is the reflectivity of the XUV multilayer spherical mirror, while the blue-white filled curve is the corrected spectrum introduced in the interaction region.

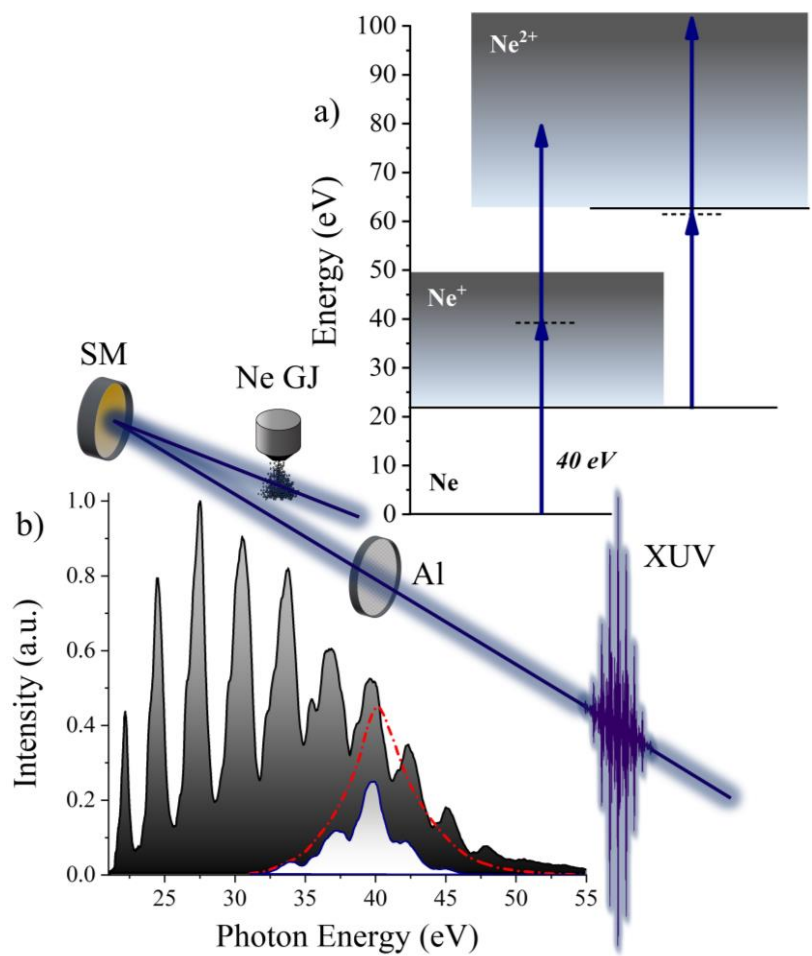

**Figure 1** (a) Two-XUV-photon double ionization scheme of Ne. Ne$^{2+}$ is formed through either a direct or a sequential ionization process, induced by an XUV pulse with 40 eV central photon energy. (b) A typical XUV spectrum. The black filled curve depicts the measured harmonic radiation, the red-dashed curve the multilayer mirror reflectivity centered at 40 eV, while the blue curve filled with white color is the corrected spectrum after the reflection by the multilayer mirror.

## Experimental results

Figure 2(a) presents a characteristic TOF mass spectrum produced by the interaction of the focused 40 eV XUV radiation with Ne gas. In the recorded spectrum singly and doubly charged Ne is observable at m/q=20 and m/q=10 respectively. The other ionic mass peaks can be attributed to rest gas and contamination species of the chamber and of the gas valve. Those include He$^+$, H$^+$, N$^+$, O$^+$, OH$^+$ and H$_2$O$^+$. It should be noted that these ions are produced by single photon ionization processes since the photoexcitation energy is exceeding their ionization energies. The inset of Figure 2(a) shows an expanded view of the portion of TOF spectrum, for clarity reasons, in which $^{20}$Ne$^{2+}$ is visible along with $^{22}$Ne$^{2+}$. The production of the doubly charged Ne species induced by a single XUV photon process would require photon energies larger than 62.5 eV. However, the spectrum reflected by the Si plate spans up to ~ 50 eV. Moreover, as mentioned, after the generation the filtered XUV beam is focused by means of a multilayer mirror with its reflectivity centered at 40 eV. Thus the observation of $^{20}$Ne$^{2+}$ is a clear evidence of a nonlinear process induced by the intense XUV radiation. In order to confirm the nonlinear nature of the process, however, two additional measurements were conducted. According to lowest-order perturbation theory one expects the ion yield produced via a non-linear process to scale proportionally with $I^n$, $I$ being the intensity of radiation and $n$ the minimum number of the photons required by the ionization process [36]. The dependence of the Ne$^{2+}$ yield on the XUV intensity has been measured and the results are presented in Figure 2(b). The measured quadratic behavior is indicative of a second order process.

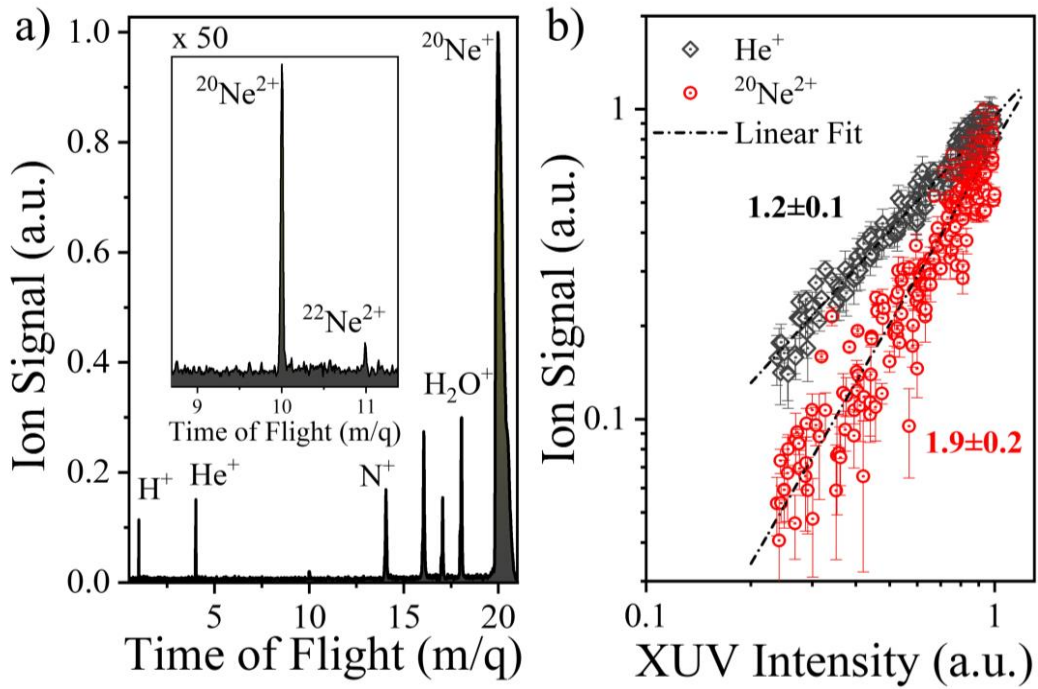

**Figure 2** (a) TOF ion mass spectrum produced by the interaction of the XUV radiation with Ne gas. (b) XUV intensity dependence of the $^{20}Ne^{2+}$ yield (red-dotted circles). The slope of 1.9±0.2 is evidencing the second order nonlinearity of the ionization process. In contrast, the $He^+$ ion yield (grey-dotted rhombus) has a linear dependence on the XUV intensity, with a slope of 1.2±0.1, as expected for a single-photon ionization process. In both cases the error bars correspond to one standard deviation of the integrated ion signals.

For the XUV intensity dependence of the double ionization of Ne, different TOF spectra were recorded and the $^{20}Ne^{2+}$ ion mass peak was integrated. The gas pressure in the interaction region was kept low, adjusted and monitored avoiding any unwanted space charge effects which affect the ionic distributions. The XUV pulse energy was varied by changing the delay between the opening of the HHG gas-jet nozzle and the arrival time of the driving IR pulse, thus changing the effective atomic density within the time interval of the interaction between the IR pulse and the gas medium. This controllable variation of the total XUV energy is illustrated by the variation of the yields of the single photon ionization products, such as the $H_2O^+$ and $He^+$ ions. The measured yield of these products is proportional to the XUV intensity in the interaction region.

As a further proof of the nonlinearity of the double ionization process, the XUV focusing mirror was translated along the propagation axis in steps of 100 μm. As demonstrated in Figure 3(a) the $^{20}Ne^{2+}$ yield is strongly decreasing when moving the focusing mirror keeping the gas jet position and the TOF's entrance fixed. On the contrary, as depicted in Figure 3(b), the products of a single photon ionization process, such as the singly charged He, show an almost constant behavior. TOF spectra were recorded at various positions of the mirror by averaging 100 shots and the ion signal was integrated. At this point it should be noted that taking into account the TOF's entrance diameter of 5 mm, the displacement range of the measurement is not expected to affect dramatically the amount of ions that enter the drift region. This is confirmed by measuring an almost flat trace in the case of the single photon ionization yields. Finally, an insignificant reduction is also expected due to the gas pressure distribution in the interaction region considering the 0.5 mm gas jet orifice and the Lorentzian expansion of the gas density, which is illustrated by the negligible variation of the $He^+$ signal. These two intensity dependent measurements are clearly evidencing the nonlinearity of the phenomena involved in the production of doubly charged Ne species.

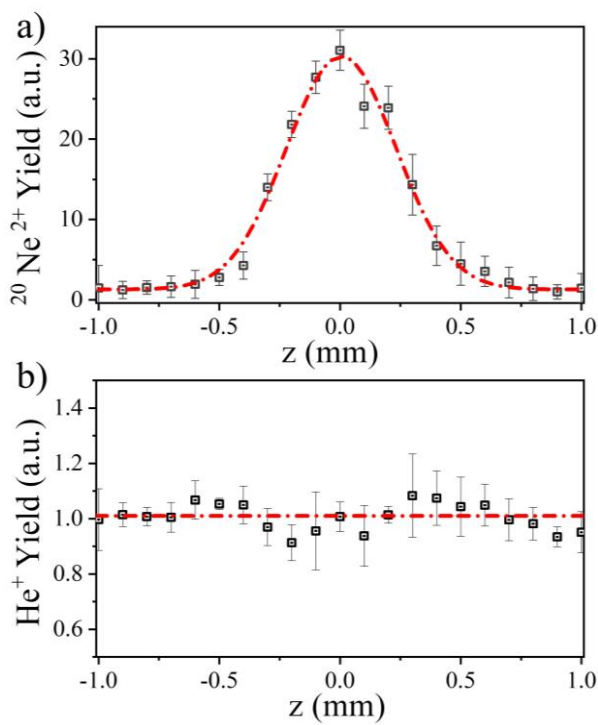

**Figure 3** (a) Measured $^{20}Ne^{2+}$ yield as a function of the position of the focus with respect to the TOF entrance. (b) $He^+$ signal as function of the focusing position. The squares are representing the measured data points with the corresponding error bars being a standard deviation of the integrated ion signal. The red lines represent a guassian (a) and a constant (b) fit of the experimental data.

**Theoretical calculations**

The interaction of the XUV radiation with Ne is treated via a second-order time dependent perturbation theory equations-of-motion (EOMs) which include the direct double ionization (DDI) and sequential double ionization (SDI) channels. In the present calculations the excitation path leading to the nearest autoionizing state (AIS) Ne→Ne($2s^{-1}3p$) at ~ 45.1 eV above the Ne ground state is also included. We have calculated its contribution to the populations and found to play an insignificant role in the doubly ionized signal relative to the direct and sequential ionization paths in accordance to similar findings in the case of helium [53]. This is due to its relative long lifetime ~ 51 fs as well as due to its large radial extension, resulting to a small electric transition amplitude for an excitation via further photon absorption. In practice this means that the relative phases of the field play no role in the results, and it is only the relative amplitudes of the individual harmonics that enter in the final equations via the pulse envelope.

Based on this observation we have omitted the AIS in the following discussion and the EOMs reduced to the below equation set:

$$\dot{\rho}_0(t) = -\left[\sum_q \gamma_{1q} I_q(t) + \sum_q \gamma_{2q} I_q^2(t)\right]\rho_0 \quad (1a)$$

$$\dot{\rho}_1(t) = -\left[\sum_{q>27} \gamma_{1q}^+ I_q(t)\right]\rho_1(t) + \left[\sum_q \gamma_{1q} I_q(t)\right]\rho_0 \quad (1b)$$

$$\dot{\rho}_{2d}(t) = \left[\sum_q \gamma_{2q} I_q^2(t)\right]\rho_0 \quad (1c)$$

$$\dot{\rho}_{2s}(\varepsilon, t) = \left[\sum_{q>27} L_q(\varepsilon)\gamma_{1q}^+ I_q(t)\right]\rho_1(t) \, , \, L_q(\varepsilon) = \frac{\gamma_{1q}^+/2\pi}{(E_1+\omega_q-\varepsilon)^2+(\gamma_{1q}^+/2)^2} \quad (1d)$$

where $I_q(t)$ represents the harmonic envelopes, $Iq \sim E_q^2(t)$. The $\rho$'s represent the populations of the different ionization stages of neon with initial conditions $\rho_0(0) = 1$ and $\rho_1(0) = \rho_{2d}(0) = \rho_{2s}(0) = 0$; $\rho_{2d}(t)$ represents the Ne$^{2+}$ population due to the DDI while $\rho_{2s}(t)$ the corresponding population to SDI pathway. The DDI path from the Ne ground state $|g\rangle$ to Ne$^{2+}$ is modelled via the 2-photon ionization rates $\gamma_{2q}^{(2)} =$

$2\pi\left|\mu_{2q}^{(0)}\right|^2$ with $\mu_{2q}^{(d)}$ being a pre-calculated two-photon dipole matrix element for the DDI channel, evaluated at $E_{2d;q} = E_0 + 2\omega_q$. For the SDI pathway, that is from the neutral Ne to the singly-ionized Neon, Ne$^+$, and from there to the doubly-ionized Ne$^{2+}$ we have calculated for the first step the partial ionization widths $\gamma_{1q} = 2\pi\left|\mu_{1q}^{(0)}\right|^2$ where $\mu_{1q}^{(0)}$ are the dipole transition matrix element from the neutral Ne to Ne$^+$ calculated at the energies $E_q = E_1 + \omega_q$, where $E_0$, $E_1$ are the energies of the neutral, Ne and the singly-charge neon, Ne+, relative the double ionization threshold; so the energy of Ne$^{2+}$ ground state is set to zero, $E_2 = 0$. For the second step the partial ionization rates are $\gamma_{1q}^+ = 2\pi\left|\mu_{1q}^{(1)}\right|^2$ where $\mu_{1q}^{(1)}$ are the dipole transition matrix element from Ne$^+$ to Ne$^{2+}$ evaluated at $E_{2s;q} = E_1 + \omega_q$. For sufficiently long pulses within a long pulse perturbation theory only the 29$^{th}$ and the 31$^{st}$ harmonics are contributing to the second step as $E_1 + \omega_q < E_2 = 0$ for q = 23, 25, 27. However the finite duration of the pulse allows contribution due to its frequency spectrum outside these sharp harmonic frequencies. This necessitated involving energy-resolved Ne$^{2+}$ states which include the contribution of these closed channels at lower frequency via an effective Lorentzian factor $L_q(\varepsilon)$ [54,55]. At the end of the pulse the SDI contribution is summed over all the respective Ne$^{2+}$ energies and the total DI signal is calculated as:

$$\rho_2 = \rho_{2d}(t \gg \tau_p) + \int d\varepsilon\, \rho_{2s}(\varepsilon, t \gg \tau_p) \qquad (2)$$

For our calculations we have used the single ionization cross section from Ne to Ne$^+$ to be, $\sigma_1 = 8\times10^{-18}$ cm$^2$ and from Ne$^+$ to Ne$^{2+}$ $\sigma_1^+ = 6.0\times10^{-18}$ cm$^2$. Our calculations showed that the cross section are relatively constant in magnitude across the pulse's frequency range. The two-photon direct double ionization cross section was calculated using uncorrelated final continuum states for the Ne$^{2+}$; the estimated value used was ~ $\sigma_{2d} = 1.4\times10^{-50}$ cm$^4$s. Unfortunately, we are lacking of a more elaborate approach to calculate this cross section due to the neon's multielectron nature which makes it a non-trivial matter to calculate doubly ionized states; nevertheless, we believe that the uncorrelated assumption provides a good order of magnitude estimation for our purposes. From the above cross sections we calculated the corresponding dipole matrix elements, $\mu_{nq}$, using the well-known relation $\sigma_n^{(j)} = 2\pi(2\pi/c)^n \omega_q^n \left|\mu_{nq}^{(j)}\right|^2$ (in a.u.) for n = 1,2 and j = 0,1 for ionization from the Ne and Ne$^+$.

Because the duration of the XUV pulse has not been measured we investigate three distinct cases. Those are: i) using an XUV pulse resulting from a Fourier Transform (FT) of the measured XUV spectrum. Although it is well known that harmonic pulses are chirped (both the envelope of the train and the individual pulses) the calculation with a Fourier Transform Limited (FTL) pulse is performed for completeness particularly because pulse compression possibilities have been previously demonstrated [56]. ii) Using a 4fs long Gaussian pulse. This duration is assumed as a result of findings of previous works [32, 35] which indicate that the duration of the envelope of the XUV pulse train is $\sim \sqrt{6}$ suggesting that the generation of plateau harmonics can be essentially treated as a $\sim 6^{th}$ order non-linear process. For the 10fs long driving laser field of the present work this corresponds to a $\sim 4$ fs long envelope. iii) Using a 3fs long Gaussian pulse. This is derived from case i), where the pulse train is essentially reduced to a single pulse (inset of Figure 4 (a)) with negligibly small side pulses. From previous works [22], [32] we anticipate that the duration of the pulses in an XUV pulse train, measured through 2$^{nd}$ order intensity volume autocorrelation, is typically about a factor of 2 larger than the FTL duration. For the present conditions this would result in an XUV duration of 3 fs.

As part of our investigation we solved the EOMs of equation (2) with pulses of peak intensity ranging from $1\times10^{12}$ to $1\times10^{13}$ W/cm$^2$ for all the three cases. The calculated Ne$^{2+}$ ion yield (in arbitrary units) resulting from the sequential and/or direct channel as a function of the XUV intensity are shown in Figure 4. The insets are the used XUV pulses. Both DDI and SDI yields are scaled over this intensity range, as presented in Figure 4, with the slope of $\approx 2$, thus confirming the presence of a second order non-linearity.

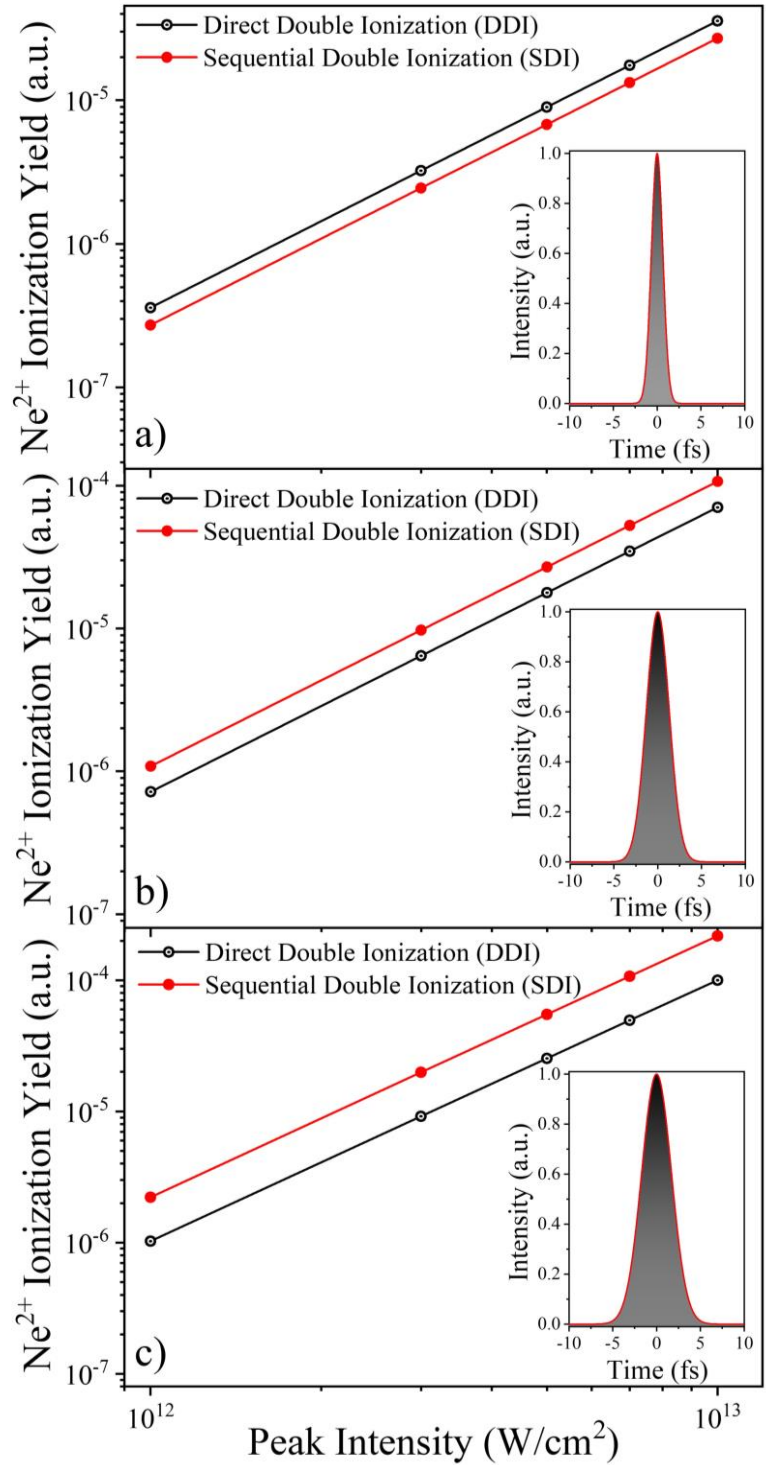

**Figure 4**: Log-log plot of the DDI and SDI yields (in arbitrary units) resulting using an FTL (a), a 3 fs long Gaussian (b) and a 4 fs long Gaussian XUV pulse (c). The slope of all lines is 2 indicative of a second order process. The ratio of the DDI to the SDI yield scales essentially linearly with the XUV pulse duration.

As depicted in Figure 4, for the excitation pulses with the given properties (peak intensity, duration) both the SDI and the DDI pathways are significantly contributing to the production of doubly charged Ne, in all three cases. While for the FTL pulse the yield of the direct process is slightly higher than that of the sequential one, in the other two cases the sequential process becomes more probable than the direct one. Since the sequential process scales with the square of the XUV pulse duration, while the direct scales linearly with it, it is expected that the ratio of the direct to the sequential process yield scales linearly with the XUV pulse duration [38]. This conclusion is in agreement with the results of our calculations.

Finally, it should be noted that the dominance of the DDI may also occur for longer pulses at higher intensities, at which the first step of the SDI is saturated and thus the SDI becomes linear with the intensity while the DDI is quadratic. However, such XUV intensities are not reached in the present work.

**Conclusions and perspectives**

In conclusion in this work two-XUV-photon double ionization of Ne induced by intense coherent XUV radiation at ELI-ALPS is demonstrated. We compared the experimental data to theoretical calculations performed using three different pulse durations. The theoretical calculations indicate that both the direct and the sequential processes are contributing to this two-XUV-photon double ionization, with their relative strength, as expected, varying linearly with the XUV pulse duration. In the present studies, the harmonic radiation is generated in Ar spanning in the range of 16-50 eV with total pulse energy of ~ 1μJ. Using high reflectivity XUV optics, a narrowband XUV spectrum with central photon energy at 40 eV and a total bandwidth of 4 eV is selected for the excitation of Ne with intensities of the order of $10^{12}$ W/cm$^2$ in the interaction region. The measurements of the Ne$^{2+}$ yield as a function of the XUV intensity revealed a non-linear behavior with a slope, in log-log scale, of nearly two, indicating the second order nonlinearity underlying the double ionization process. Second-order time dependent perturbation theory equations-of-motion indicate the dominance of the direct double ionization pathway. The calculated ion yields are scaling over the intensity range of $1\times10^{12}$ - $1\times10^{13}$ W/cm$^2$ with the slope of ≈ 2, in agreement with the second order non-linearity measured experimentally. The presented results establish the recently constructed SYLOS GHHG-COMPACT beamline at ELI-

ALPS as an intense attosecond beamline with which nonlinear XUV processes can be studied and/or used. A near future perspective is to drive the beamline with the now operational 1kHz repetition rate laser chain at ELI-ALPS. Two-XUV-photon direct double ionization at 1 kHz repetition rate enables electron-electron and electron-ion coincidence experiments which will shed light onto several open questions in electron-electron correlation and other ultrafast processes. The 1kHz SYLOS GHHG-COMPACT beamline will host an advanced Reaction Microscope end-station dedicated to such investigations.

**Acknowledgments**

We acknowledge support of this work by the Hellenic Foundation for Research and Innovation (HFRI) and the General Secretariat for Research and Technology (GSRT) under grant agreements [GAICPEU (Grant No 645)] and NEA-APS HFRI-FM17-3173, the H2020 project IMPULSE (GA 871161). The ELI-ALPS Project (GINOP-2.3.6-15-2015-00001) is supported by the European Union and it is co-financed by the European Regional Development Fund. S.K. acknowledges Project No. 2020-1.2.4-TÉT-IPARI-2021-00018, which has been implemented with support provided by the National Research, Development and Innovation Fund of Hungary, and financed under the 2020-1.2.4-TET-IPARI-CN funding scheme. H.S., B.B., and M.F.K. are grateful for support by the Max Planck Society within the framework of the Max Planck Fellow program. M.F.K.'s work at SLAC is supported by the U.S. Department of Energy, Office of Science, Basic Energy Sciences, Scientific User Facilities Division. G.S. acknowledges financial support by the Deutsche Forschungsgemeinschaft grant 429805582 and by the Bundesministerium für Bildung und Forschung (Project 05K19VF1). L.A.A.N. and A.F. also acknowledge the support of the Irish Research Council under the Govt PG Scholarship (GOIPG/2018/1070). I.O. and E.S. acknowledge support by the ELI-ALPS team during the experimental runs.